\documentclass[aps,twocolumn,superscriptaddress,showpacs]{revtex4}
\usepackage{amsmath,amssymb,graphicx}
\usepackage{color}

\graphicspath{{fig-ps/}}
\graphicspath{{figures/}}

\begin{document}

\title{Single and Multiple Vortex Rings in Three-Dimensional Bose-Einstein Condensates:
Existence, Stability and Dynamics}

\author{Wenlong Wang}
\email{wenlongcmp@gmail.com}
\affiliation{Department of Physics and Astronomy, Texas A$\&$M University,
College Station, Texas 77843-4242, USA}

\author{R.N.~Bisset}
\affiliation{INO-CNR BEC Center and Dipartimento di Fisica,
Universit\`a di Trento, Via Sommarive 14, I-38123 Povo, Italy}

\author{C.~Ticknor}
\affiliation{Theoretical Division, Los Alamos
National Laboratory, Los Alamos, NM 87545}

\author{R.~Carretero-Gonz{\'a}lez}
\affiliation{Nonlinear Dynamical Systems
Group,\footnote{URL: \texttt{http:/\!/nlds.sdsu.edu}}
Computational Sciences Research Center, and
Department of Mathematics and Statistics,
San Diego State University, San Diego, California 92182-7720, USA}

\author{D.J.~Frantzeskakis}
\affiliation{Department of Physics, University of Athens,
Panepistimiopolis, Zografos, Athens 15784, Greece}

\author{L.A.~Collins}
\affiliation{Theoretical Division, Los Alamos
National Laboratory, Los Alamos, NM 87545}

\author{P.G.~Kevrekidis}
\email{kevrekid@math.umass.edu}
\affiliation{Department of Mathematics and Statistics, University of Massachusetts,
Amherst, Massachusetts 01003-4515 USA}

\begin{abstract}
In the present work, we explore the existence, stability and
dynamics of single and multiple vortex
ring states that can arise in Bose-Einstein condensates.
Earlier works have illustrated the bifurcation of such states,
in the vicinity of the linear limit, for isotropic or anisotropic
three-dimensional harmonic traps. Here, we extend
these states to the regime of large chemical potentials, the so-called
Thomas-Fermi limit, and explore their properties such as equilibrium radii 
and inter-ring distance, for multi-ring states, as well as their vibrational spectra and possible
instabilities. In this limit, both the existence and stability characteristics
can be partially traced to a particle picture that considers the
rings as individual particles oscillating within
the trap and interacting pairwise with one another.
Finally, we examine some representative instability scenarios of 
the multi-ring dynamics including breakup and reconnections, 
as well as the transient formation of vortex lines.
\end{abstract}

\pacs{67.85.-d, 67.85.Bc, 47.32.cf, 03.75.-b}

\maketitle

\section{Introduction}

Bose-Einstein condensates (BECs) of ultracold atomic gases have for
around two decades now~\cite{book1,book2,dumitr1,dumitr2} captured the interest
not only of the atomic, molecular, and optical physics communities, 
but also considerably so that
of the nonlinear waves~\cite{emergent,book_new}.
This, to a significant degree, is due to the effective nonlinearity
introduced at the lowest-order mean-field theory~\cite{book1,book2,book_new}, which leads to   
a nonlinear Schr{\"o}dinger
type equation~\cite{abl1,abl2,abl3,abl4}, 
referred to as the Gross-Pitaevskii equation (GPE). Depending on the sign of the $s$-wave scattering
length, this results in a self-defocusing
or self-focusing nonlinearity in the equation
which, in turn,
admits an array of relevant nonlinear wave structures. 
The latter comprise, but are not limited to,
bright~\cite{expb1,expb2,expb3}, gap~\cite{gap} and
dark~\cite{djf}  matter-wave solitons.
In higher dimensions the main structures are
vortices~\cite{fetter1,fetter2}, in two dimensions, as well as 
vortex lines 
and rings, in three dimensions~\cite{komineas_rev}.

Our focus in the present work will be based on vortex
rings (VRs) arising in three-dimensional (3D) repulsive BECs. Such
structures have been studied rather extensively in
theoretical, computational, and experimental works,
with a number of reviews having emerged from this
activity~\cite{emergent,komineas_rev,book_new}.
Despite the wide body of knowledge
existing from the physics of fluids and superfluids~\cite{saffman,Pismen1999} 
and the much earlier experimental observation of 
these structures 
in helium~\cite{donnelly,Rayfield64,Gamota73}, 
the amount of experimental attention that VRs have received
in BECs is less than might be expected given their importance.
Nevertheless, they have been experimentally observed in a range of different
studies including, their realization through 
the decay of planar dark solitons in two-component BECs~\cite{Anderson01},
creation via density engineering~\cite{Shomroni09}, spontaneous
emergence via the collision of symmetric defects~\cite{Ginsberg05}
and their detection through unusual collision outcomes of dark 
solitonic structures~\cite{sengstock}. Recently, intriguing 
six-vortex-ring and six-vortex-line cages have been predicted 
as transient states in the decay of spherical-shell dark solitons~\cite{dsnew}.

In recent years, a program of exploring coherent structures in a two-pronged 
way has naturally emerged and has been summarized, e.g.,~in Ref.~\cite{book_new}.
On the one hand, it is possible to study 
nonlinear waves in the vicinity of the linear (noninteracting) limit of
the GPE, namely that of the quantum harmonic oscillator.
This limit, while physically less relevant given its association with small
atom numbers, is very insightful towards the waveforms that are possible via different
combinations of the linear eigenfunctions; incidentally, this regime is 
also more prone to important quantum-fluctuation phenomena~\cite{prouk}.
In the context of VRs and related states, this approach has been utilized, e.g.,
in Refs.~\cite{danaila,ourpra1} to construct single and multiple VR 
structures.
On the other hand, a complementary approach that is
certainly relevant experimentally is the exploration of the highly
nonlinear, large atom number regime.
This is known as the Thomas-Fermi (TF) limit, in which the structures become narrower 
as the healing length, which constitutes their characteristic length scale, decreases.
Here, the coherent waves can be thought
of as individual ``particles'' that have kinematics and inter-particle
dynamics that can be approximated by suitable particle models.
Although attempts have been made to explore a single VR
as such a particle, to describe its vibrational
modes~\cite{jackson99,fetterpra,horng,ourpra2}, as well
as to explore multiple VRs in a homogeneous
(untrapped) medium~\cite{konstantinov,Shashikanth03,caplan}, to the best
of our knowledge no such attempt has been made
for the case of multiple trapped VRs. It is the purpose of this work to 
contribute to this direction by developing a systematic approach to study this problem. 
We thus show that multiple trapped VRs are 
especially important: this is due to the fact that 
the interplay of the intrinsic dynamics of each VR, with the trapping 
and inter-VR interactions, produce the possibility of stationary multi-VR
states that we will argue ---based on their stability properties--- 
are experimentally accessible. It is also shown that 
multiple VRs also produce intriguing vibrational dynamics, within both
the stable and unstable regimes.

The layout of the paper is as follows. In Sec.~\ref{Sec:GPE} we outline the numerical formulation
of the three-dimensional GPE and its Bogoliubov-de Gennes (BdG) spectral stability analysis.
%
In the theoretical formulation of Sec.~\ref{sec:sub:part},
we present the ``particle picture'' (PP) that we will use to provide 
insights into the single and multiple VR steady states.
Then, in Sec.~\ref{results}, we discuss our numerical results; firstly,
for the cases of the single and double VRs, and finally for
that of a triple VR state that we can also systematically
construct and probe. We make direct comparisons between the treatments 
to illustrate the qualitative ability of the theory to capture multi-VR states.
Our existence and stability computations, together with the PP
analysis, are complemented with a selection of dynamical
manifestations of both the vibrational modes ---such as those 
that describe the relative motion of multiple VRs--- and 
the intriguing instabilities in certain regimes.
Finally, we summarize our findings, discuss some open
problems and present a number of
possibilities for future studies in Sec.~\ref{conclusion}.

\section{Theoretical and Computational Model Setup}

\subsection{The Gross-Pitaevskii equation}\label{Sec:GPE}

In the framework of the lowest-order mean-field theory, for sufficiently
low temperatures, the dynamics of a 3D repulsive BEC is well-described by the
GPE~\cite{book1,book2,emergent,book_new},
\begin{equation}
i\hbar \psi_t=-\frac{\hbar^2}{2m} \nabla^2 \psi+V \psi + g| \psi |^2 \psi-\mu \psi,
\label{GPE}
\end{equation}
where $\psi(x,y,z,t)$ is the macroscopic wavefunction,
$\mu$ is the chemical potential and $g=4\pi\hbar^2a_s/m$, with 
$a_s$ being the $s$-wave scattering length and $m$ the atomic mass. 
We use a harmonic trap of the form 
\begin{equation}
V(\vec{r})=\frac{1}{2} m(\omega_{r}^2 r^2
+\omega_z^2 z^2),
\label{potential}
\end{equation}
where $r^2=x^2+y^2$, while $\omega_{r}$ and $\omega_z$ are the
trapping frequencies along the radial and
axial directions, respectively.
Note that the potential has rotational symmetry with respect to the $z$-axis.
In our simulations we study the
two-VR state with $\omega_z=\omega_r$, while we explore
the three-VR state with $\omega_z=\frac{2}{3}\omega_r$. In these settings, the two states have chemical potentials at the linear (non-interacting)
limit equal to $\mu_c=\frac{7}{2}\hbar\omega_r$ and $\frac{10}{3}\hbar\omega_r$,
respectively.
This occurs when the ring-dark-soliton state
becomes energetically degenerate with either the two
or three-planar-dark-soliton state, 
enabling combinations of two states with a relative phase difference of $\pi/2$ 
that give rise to the two-VR or three-VR state, respectively~\cite{danaila,ourpra1}.
These states can then be numerically followed all the way from the above
mentioned linear limit to the large chemical potential, highly
nonlinear limit. We have also performed the corresponding BdG 
spectral analysis of the states as a function of the chemical potential using the linearization ansatz, 
\begin{eqnarray}
\psi(\vec{r},t)= \psi_0(\vec{r}) + \epsilon \left( a(\vec{r}) e^{\lambda t}
+ b^*(\vec{r}) e^{\lambda^{*} t} \right).
\label{bdg}
\end{eqnarray}
Here, $\psi_0$ denotes the single- or multi-VR stationary state whose
stability is sought,
$\epsilon$ is a formal perturbation parameter and $\lambda$ denotes
the eigenvalue corresponding to the eigenvector $(a,b)^T$, with $(\cdot)^T$
denoting the transpose and $(\cdot)^*$ complex conjugation.

We temporally evolve the GPE, Eq.~(\ref{GPE}), using two independent methods, 
providing a strong test of our numerics. 
In one method, we utilize a real-space product scheme with a 
finite-element discrete-variable representation. 
This is based on a split-operator approach, and a Gauss-Legendre quadrature is 
implemented within each element~\cite{Schneider2006,dvr-exp}.
In the other method, a split-step operator is performed on a Fast-Fourier-Transform (FFT) grid.
Our BdG calculations are made feasible by implementing a Fourier-Hankel scheme 
that utilizes azimuthal symmetry, much as was done in Refs.~\cite{Ronen2006,Blakie2012}.
In addition to allowing one to treat 3D functions as 2D, numerically, this allows 
the diagonalization of each angular momentum subspace individually.

In what follows, we are interested in both the properties 
of the GPE solutions themselves, such as the equilibrium radial or axial positions, and those
of their excitations. We will also attempt to connect the GPE solutions with our particle picture results.

\subsection{The particle picture of vortex rings in a trap}
\label{sec:sub:part}


We now focus on the limit of large chemical potentials $\mu$,
where we can provide a theoretical analysis of the in-trap dynamics
and interactions of multiple VRs. 
For simplicity, in 
all that follows
we have used dimensionless units (see,
e.g., Refs.~\cite{emergent,book_new}) where time is measured in units
of inverse trap frequency ($1/\omega_r$), length scales in units of 
harmonic oscillator length ($a_r = \sqrt{\hbar/m \omega_r}$) and energy units of $\hbar\omega_r$.
In this TF limit, there exists a well known
approximation to the
ground state of the GPE given by
$\psi_{\text{TF}}=\sqrt{\max[\mu-V(\vec{r}),0]}$.
Here, our aim is to explore vortical excitations (in particular,
VRs) on top of this ground state.
As indicated in the introduction, studies have independently considered 
each of the following:
\begin{itemize}
\item[(i)] the motion of a single VR in a homogeneous medium~\cite{roberts},
\item[(ii)] the effect of  a trap on a single VR~\cite{JacksonMcCannAdams,jackson99,fetterpra,horng}, and the
\item[(iii)] interaction of multiple VRs in the absence of a trap~\cite{konstantinov,Shashikanth03,caplan}.
\end{itemize}
{%
Another aim of this work, in addition to exploring the
existence, stability, and dynamics of these states, is to
explore the effective PP model arising from combining
these different ingredients together, and its usefulness in
capturing the essential static properties corresponding
to equilibrium configurations of multiple VRs in a trap
as well as the periodic oscillations ensuing from 
initial conditions away from these equilibria.}

For a set of co-axial VRs along the $z$-axis, 
a na\"{\i}ve approach to combine the above-mentioned VR-VR and VR-trap 
contributions would consist of simply {\em adding} the corresponding
reduced dynamics at the level of the effective ordinary differential 
equations (ODEs) on the VR radii $r_i$ and positions $z_i$.
However, perhaps somewhat surprisingly, this approach turns out to produce
{\em non-Hamiltonian} ODEs because the two
main contributions, namely the VR-VR interaction~\cite{konstantinov} 
and VR-trap interaction~\cite{JacksonMcCannAdams}, originate from
energy terms with {\em different} canonical variables (see below).
Therefore, this approach ---although capable of reasonably predicting the 
positions for stationary multi-VR configuration (results not shown here)---
fails to describe the actual dynamics of trapped multi-VR configurations.
In fact, the ensuing VR dynamics for this na\"{\i}ve approach
give rise to damped/anti-damped orbits (depending on the VR charges) 
that tend to spuriously spiral in/out of the condensate.

{%
In order to derive a self-consistent, Hamiltonian set of equations for trapped multi-VR
configurations it is necessary to start from the Hamiltonian formulation
of the different interaction energies involved and then obtain the
equations of motion through a common set of canonical variables.
Let us then consider the following energies:
(i) the VR-trap energy, denoted by $E_{\text{VR-T}}$, described in Ref.~\cite{JacksonMcCannAdams}
and (ii) the VR-VR energy, denoted by $E_{\text{VR-VR}}$, described in Ref.~\cite{konstantinov}.
Importantly, both $E_{\text{VR-T}}$ and $E_{\text{VR-VR}}$ contain
the VR self-induced velocity that is responsible for a single VR to always
have an intrinsic velocity. Therefore, we construct the total energy of the
system with the following combination that only includes the self-induced
interaction once:
\begin{eqnarray}
\notag
E &=& 
E_{\text{VR-T}} + E_{\text{VR-VR}} - E_{\text{VR-VR}}^{\text{self}},
\\[1.0ex]
  &=& 
E_{\text{VR-T}} + \tilde{E}_{\text{VR-VR}} 
\label{eq:E_TOT}
\end{eqnarray}
where $E_{\mbox{\scriptsize VR-VR}}^{\text{self}}$ corresponds to the contribution 
to the energy originating from the self-induced velocity in the untrapped 
VR-VR description~\cite{konstantinov}.
Specifically, using Refs.~\cite{JacksonMcCannAdams} and~\cite{konstantinov},
to describe these contributions yields the following energies.
The VR-T contribution for a single VR at position $(r_i,z_i)$ inside an 
isotropic ($\omega=\omega_r=\omega_z$) trapping potential with
TF radius $R_\perp=\sqrt{2\mu}/\omega$ yields~\cite{JacksonMcCannAdams}
\begin{eqnarray}
E_{\text{VR-T}} &=& 2\pi\,\mu\, r_i \times \\[1.0ex]
\notag
&& \left[ \left(1-\frac{r_0^2}{R_\perp^2} \right) \ln \left(\frac{\sqrt{R_\perp^2-r_0^2}}{\xi} \right) + \frac{r_0^2}{R_\perp^2} - 1\right],
\label{eq:E_VR_T}
\end{eqnarray}
with $r_0^2=r_i^2+z_i^2$ and where $\xi=1/\sqrt{a\mu}$ is obtained from the
$r\approx0$ asymptotics of the vortex core density $\rho(r)\approx a\mu^2 r^2$
and $a=0.82226$ was computed numerically by fitting this asymptotic
expression.
On the other hand, the VR-VR contributions, without self-induced
velocity terms, for a set of $N$ VRs of charge $m_i$ and position
$(r_i,z_i)$ yield~\cite{konstantinov}
\begin{eqnarray}
\tilde{E}_{\text{VR-VR}} = 4\pi\sum_{\substack{i,j=1 \\ i\neq j}}^N 
 m_i m_j\, \sqrt{r_i r_j}\,C(k_{ij}),
\label{eq:E_VR_VR}
\end{eqnarray}
where
\begin{eqnarray}
C(k_{ij})&=&
\left(\frac{2}{{k_{ij}}}-{k_{ij}}\right){\cal K}(k_{ij})-\frac{2}{{k_{ij}}}{\cal E}(k_{ij}),
\label{eq:Cij}
\\[1.0ex]
\label{eq:kij}
k_{ij}^2&=&\frac{4r_ir_j}{(z_i-z_j)^2+(r_i+r_j)^2}.
\end{eqnarray}
and ${\cal K}$ and ${\cal E}$ are, respectively, the complete elliptic integrals of the
first and second kind and $k_{ij}$ is their respective elliptic modulus.
}

Having the total energy~(\ref{eq:E_TOT}), i.e., the Hamiltonian
for the system of interacting VRs, we can obtain the equations of motion 
as a set of coupled ODEs for the vortex positions using the 
corresponding Hamilton's equations:
\begin{eqnarray}
\dot{p}_i=-\frac{\partial E}{\partial q_i} \quad \text{and}\quad
\dot{q}_i= \frac{\partial E}{\partial p_i}.
\label{eq:HJ}
\end{eqnarray}
We follow the choice of canonical variables $(p_i,q_i)$ 
from Ref.~\cite{konstantinov}:
\begin{eqnarray}
(p_i,q_i)=(2\pi\,m_i\,  r_i^2,z_i).
\label{eq:canonical}
\end{eqnarray}
It is precisely due to this choice of canonical variables that 
it is not possible to apply the na\"{\i}ve approach of adding
the resulting equations on motion from Refs.~\cite{konstantinov} 
and \cite{horng} as the former uses the canonical variables~(\ref{eq:canonical})
while the latter uses $(p_i,q_i)=(r_i,z_i)$.
For completeness, the resulting ODEs are included in
the Appendix.

\begin{figure}[tb]
\begin{center}
\includegraphics[width=8cm]{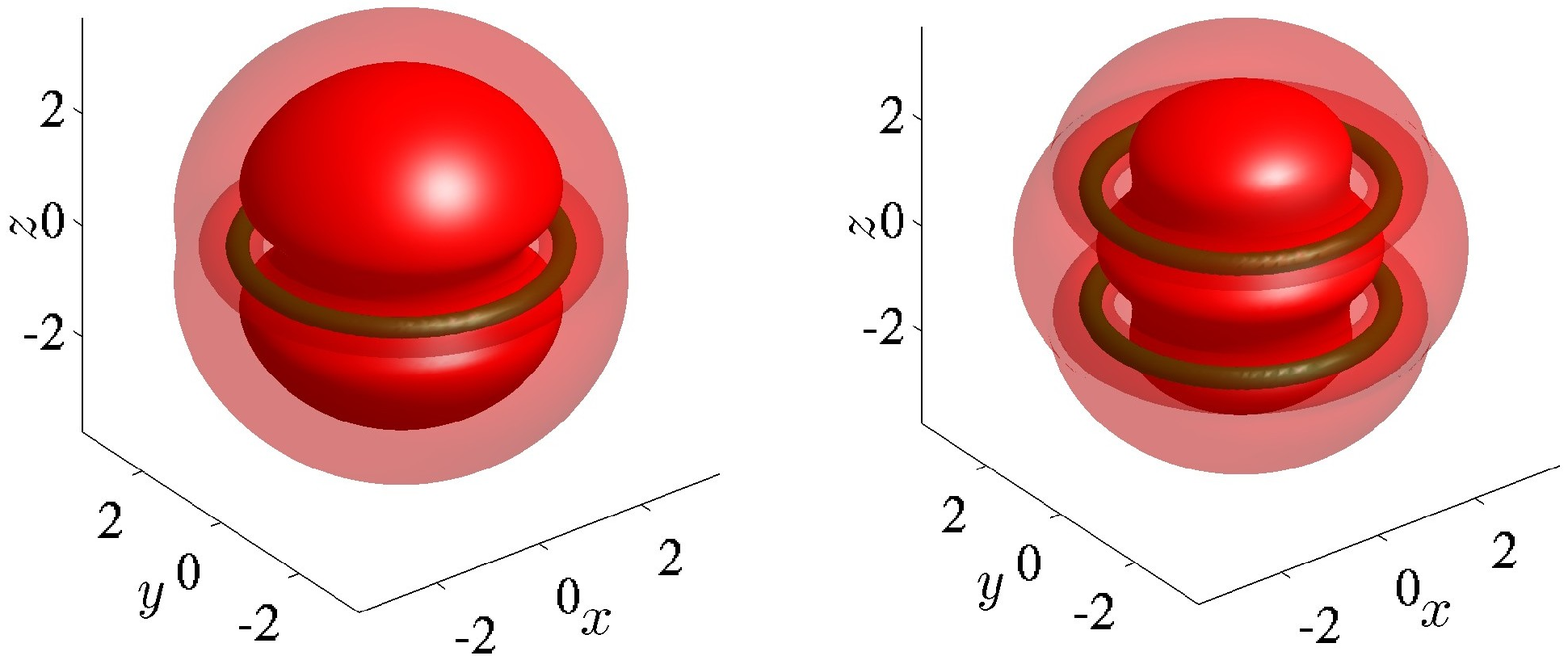}\\[1.0ex]
\includegraphics[width=4cm]{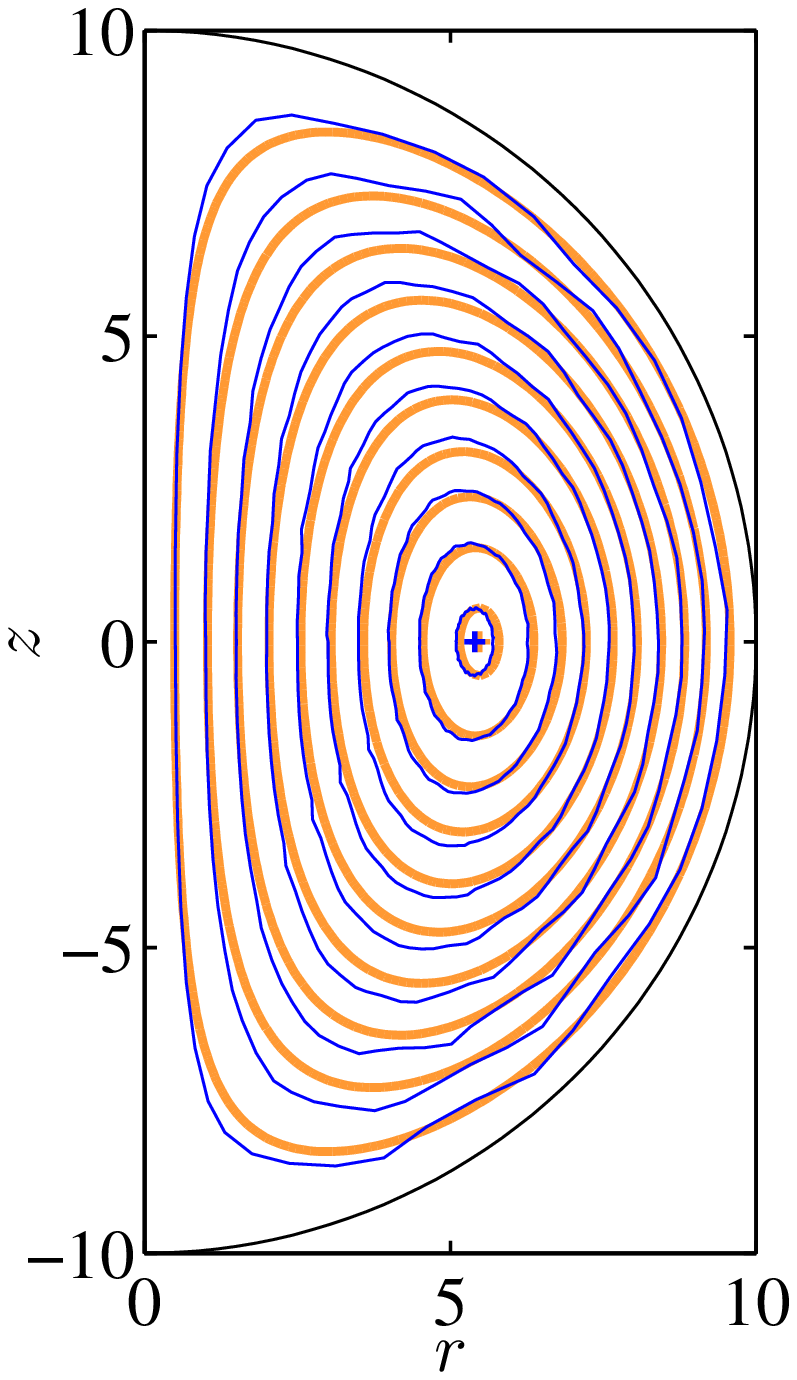}
\includegraphics[width=4cm]{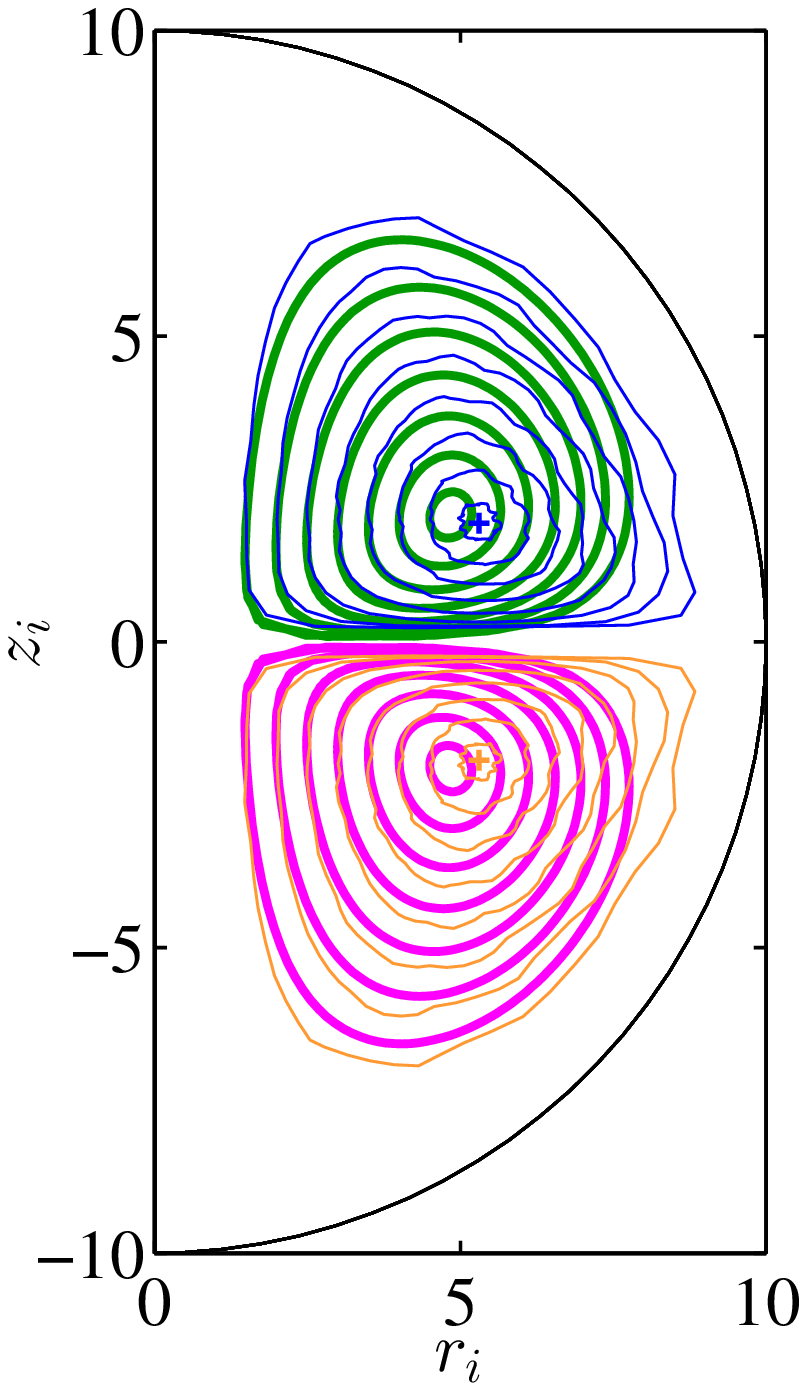}
\caption{(Color online)
{%
The top row of panels illustrates {\em stationary} single (left) 
and double (right) VRs for $\mu=12$
in an isotropic trap with $\omega_z=\omega_r=1$.
The panels depict isocontour plots for the density
(red) and the cores of the VRs are highlighted by green (dark) surfaces
corresponding to isocontours of the smoothed norm of the
vorticity (curl of the fluid velocity).
The bottom row of panels depicts the trajectories for
one VR (left) and two oppositely charged VRs (right)
in an isotropic trapping with $\omega_z=\omega_r=1$ and $\mu=50$.
Using the cylindrical symmetry of the setup, the trajectories 
are depicted in the $(r,z)$ plane where $r$ is the radius
of the VR and $z$ its vertical axis coordinate.
The thick solid trajectories correspond to the particle
picture (PP) prediction of Sec.~\ref{sec:sub:part} while
the thin solid trajectories correspond to numerical
simulations of the GPE of Eq.~(\ref{GPE}) integrated in 
reduced cylindrical coordinates. 
The outermost semi-circle corresponds to the TF radius.
For full 3D numerics showing dynamic and symmetry-breaking 
instabilities, we refer the reader to the results in Figs.~\ref{Fig:2VR_EA}, 
\ref{Fig:2VR_EB}, \ref{fig:dyn1} and \ref{fig:dyn2}.
}
}
\label{fig:RZcuts}
\end{center}
\end{figure}

\begin{figure}[tb]
\begin{center}
\includegraphics[width=8cm]{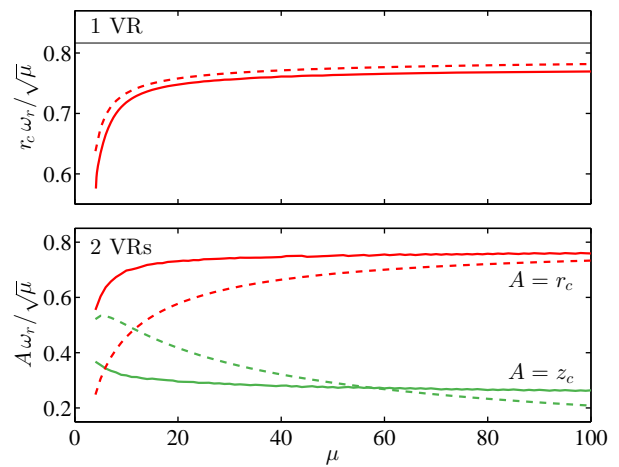}
\caption{(Color online)
%
%
{%
Equilibrium positions for one and two VR configurations
as a function of the chemical potential $\mu$.
}%
The top and bottom panels show the equilibrium radius $r_c$ and 
axial location $z_c$, as a function of the dimensionless (by $\hbar \omega_r$) 
chemical potential $\mu$, for the particle picture (PP) (dashed line) and
the GPE (solid line) for the single VR (upper panel)
and double VR (bottom panel) for $\omega_z=\omega_r=1$.
%
{%
The top panel includes (thin horizontal line) the asymptotic prediction 
for $\mu\rightarrow\infty$ given in Eq.~(\ref{rc1}).
}
All quantities in this and subsequent figures are dimensionless, see text.
}
\label{fig:SteadyStates}
\end{center}
\end{figure}

\section{Results}
\label{results}

We begin the discussion of our numerical findings by presenting existence
results for the case of single and multiple VRs.
{%
The top row of panels in Fig.~\ref{fig:RZcuts} depicts illustrative
examples of the steady-state configurations for one and two VRs.
However, if the VRs are not placed at their stationary position, they start to 
oscillate as depicted in the bottom row of panels in Fig.~\ref{fig:RZcuts}.
These panels depict the trajectories for one VR (left panel)
and two oppositely charged VRs (right panel) in an isotropic  parabolic 
trap (\ref{potential}) with $\omega_r=\omega_z=1$ and chemical potential $\mu=50$.
The thick and thin trajectories correspond, respectively, to our PP and the GPE dynamics.
The figure suggests that the dynamics for a single VR is qualitatively and 
quantitatively very well described by the PP. The case of two VRs (right panel)
indicates a good qualitative match between PP and original GPE dynamics, yet
the position of the steady state configurations seems to be slightly shifted.
}
%
%
{%
In order to have a broader sense of the validity of the PP, let us
follow the steady state VR positions for one and two VRs as the
chemical potential is varied.
Figure~\ref{fig:SteadyStates} depicts the comparison between the numerical 
solutions of the GPE (solid lines) and the PP (dashed lines). 
For a single VR (top panel), the PP does an excellent job at
predicting the stationary position of the VR and its functional
dependence on the chemical potential $\mu$.
Note that the $\mu\rightarrow\infty$ asymptotic prediction in the TF limit
from Refs.~\cite{kaper,dsnew} for the equilibrium radius $r_c$:
\begin{eqnarray}
r_c= \sqrt{\frac{2 \mu}{3 \omega_r^2}} ,
\label{rc1}
\end{eqnarray}
is slightly higher than the one predicted by the PP.
}
%

{%
The bottom panel of Fig.~\ref{fig:SteadyStates}
depicts the comparison between the GPE and PP
results for the two-VR steady state configuration.
}
In this case, one can see that the trends predicted by the PP can 
qualitatively follow that of the full GPE, although
some quantitative disparity 
remains. This can be attributed to {the following} causes: 
\begin{itemize}
\item[(i)] {the PP is an amalgamation of different contributions
  stemming from different approaches; it would be useful, although
  technically seemingly especially
  tedious to consider an approach incorporating
all three effects concurrently.}
%
\item[(ii)] The accumulation of errors for the three
  different contributions, since each one of them involves
  corresponding approximations.
\item[(iii)] the interaction between the VRs is modulated by density 
  variations, a feature that is not captured in the effective PP in the form
  considered herein.
\end{itemize}
This last point is also an issue that affects similar particle approaches
for vortices in 2D, and has been discussed in
Refs.~\cite{busch,middelkamp}. 
Nevertheless, we conclude that
the PP approach, while less quantitatively
dependable, can be used to provide a qualitative handle on the
trends of stationary multi-VR characteristics.

\begin{figure}[tb]
\begin{center}
\includegraphics[width=8.0cm]{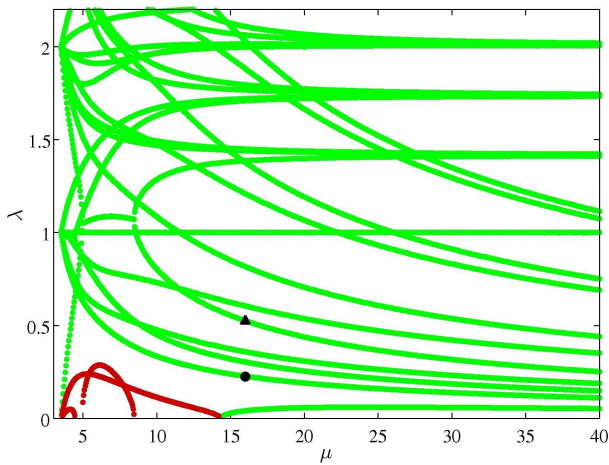}
\caption{(Color online) 
BdG spectrum for the two-VR as a function
of the dimensionless chemical potential $\mu$ for $\omega_z=\omega_r=1$.
Mode contributions are shown via green (light) points if stable and 
red (dark) points if unstable. 
The black circle and triangle symbols represent the frequencies
for the normal modes of vibration off the stationary state
depicted, respectively, in Figs.~\ref{Fig:2VR_EA} and \ref{Fig:2VR_EB}.
%
%
%
}
\label{fig:spectrum2VR}
\end{center}
\end{figure}


\begin{figure}[tb]
\begin{center}
\includegraphics[width=8cm]{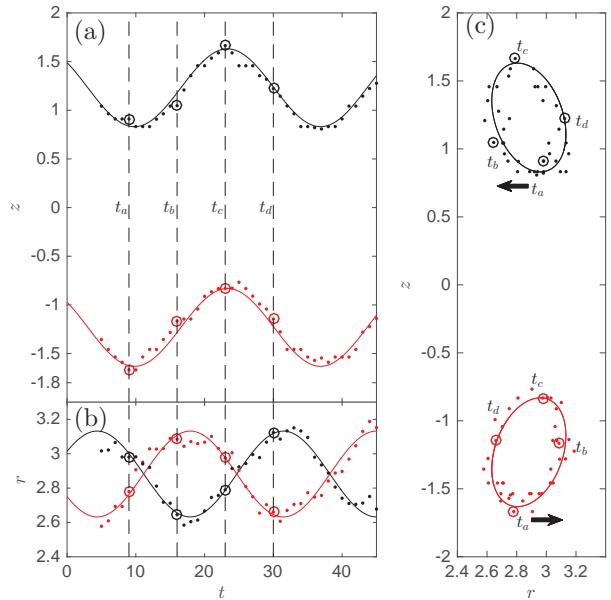}
\caption{(Color online)
The first $n=0$ excitation of the two-VR state. 
Black/red (gray) represents the upper/lower ring.
The open circles (and corresponding vertical dashed
lines) indicate four time points to demonstrate the motion of the rings via their
(a) time-$z$, (b) time-$r$ and (c) ($r,z$) trajectories.
The solid curves are fits to the data, while the arrows in (c) indicate time
$t_a$ and the subsequent direction of core motion. 
Time and space are measured in units of $1/\omega_r$ and
harmonic oscillator length $a_r = \sqrt{\hbar/m \omega_r}$, respectively.
}
\label{Fig:2VR_EA}
\end{center}
\end{figure}

\begin{figure}[tb]
\begin{center}
\includegraphics[width=8cm]{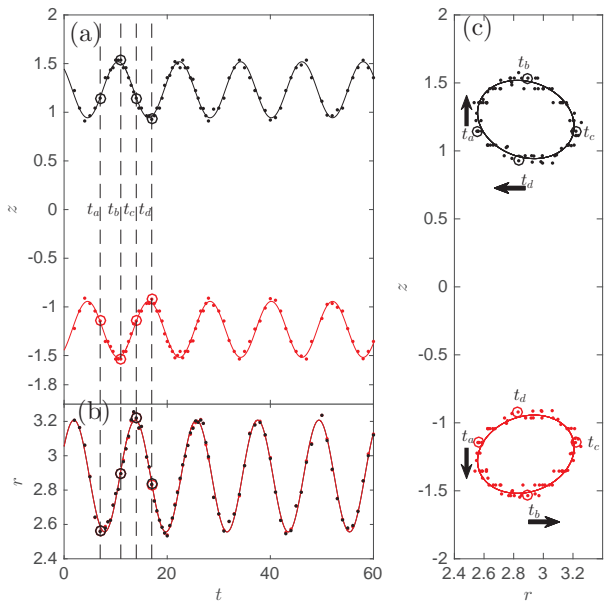}
\caption{(Color online) The second $n=0$ excitation of the two-vortex-ring state.
Curves and labels have the same meaning as in Fig.~\ref{Fig:2VR_EA}.
These rings, as was shown in Fig.~\ref{Fig:2VR_EA}  for the first excitation, 
have azimuthal symmetry and cores that remain circular and coaxial with 
the $z$-axis throughout the motion. 
Note that in (b) the $r$-motions of the two rings are in-phase and the data 
points overlap. As explained in the caption of Fig.~\ref{Fig:2VR_EA},
$\mu\sim 16$.
%
Time and space are measured in units of $1/\omega_r$ and
harmonic oscillator length $a_r = \sqrt{\hbar/m \omega_r}$.
}
\label{Fig:2VR_EB}
\end{center}
\end{figure}

We now turn to the exploration of the spectrum
of the multi-VR states. Although a brief
discussion of this can be found in Ref.~\cite{ourpra1}, in the vicinity of
the linear limit,
here we consider the relevant spectrum more systematically for
a wide parameter range.
%
%
The relevant spectrum for the two-VR solution is shown in Fig.~\ref{fig:spectrum2VR}, 
and is seen to bear numerous similarities to that of the single
VR (discussed in Ref.~\cite{ourpra2}).
In particular, the modes associated with dynamical instability are fairly 
similar to those of the single VR. The most significant one
among them, associated with the widest parameter range of instability,
is related to a quadrupolar mode having an $n=2$ azimuthal dependence
$e^{i n \theta}$, as we will also see below in the dynamical-evolution results.
On the other hand, the mode immediately above this one, for
large $\mu$ in the spectrum, is azimuthally symmetric ($n=0$) and is associated with a stable relative motion of the two vortex rings.
The trajectory of this mode is illustrated in Fig.~\ref{Fig:2VR_EA}.
This excitation was initiated by mixing the two-VR stationary state with the corresponding BdG excitation; 
the resulting wavefunction was then renormalized to preserve the total atom number and subsequently 
evolved in time according to the GPE.
Since this excitation has $n=0$,
these rings have azimuthal symmetry, and the cores remain circular and coaxial
with the $z$-axis throughout the motion. 
In the ($r,z$) cross section of Fig.~\ref{Fig:2VR_EA}(c), the upper ring 
orbits clockwise while the lower ring orbits counterclockwise in such a way that their $z$ 
motion remains in-phase.
The fitted period for this oscillation is $T= 27.1$ (in units of $1/\omega_r$), 
which compares favorably with the BdG prediction $T= 27.9$
(see black circle in Fig.~\ref{fig:spectrum2VR}).
Note that the chemical potential is not well-defined in the absence of a 
stationary state, of which was (in dimensionless units) $\mu\sim 16$
before the addition of the excitation.
%
%

Contrary to the single VR studied in Ref.~\cite{ourpra2}, the existence of an 
additional ring means that there is a second $n=0$ vortex excitation.
This is shown in Fig.~\ref{Fig:2VR_EB}, and exhibits a similar motion to 
the first excitation but now the measured period is $T= 11.9$,
compared to the BdG period of $T= 12.0$ (see black triangle symbol in
Fig.~\ref{fig:spectrum2VR}).  Furthermore, the relative phase of 
the ring motion differs such that their $r$ positions are now in-phase,
while the $z$ positions now perform an out-of-phase oscillation.
%
%
Finally, the branches that level off at large $\mu$ are the non-vortex excitations 
of the underlying ground state.


%


\begin{figure}[tb]
\begin{center}
\includegraphics[width=7.0cm]{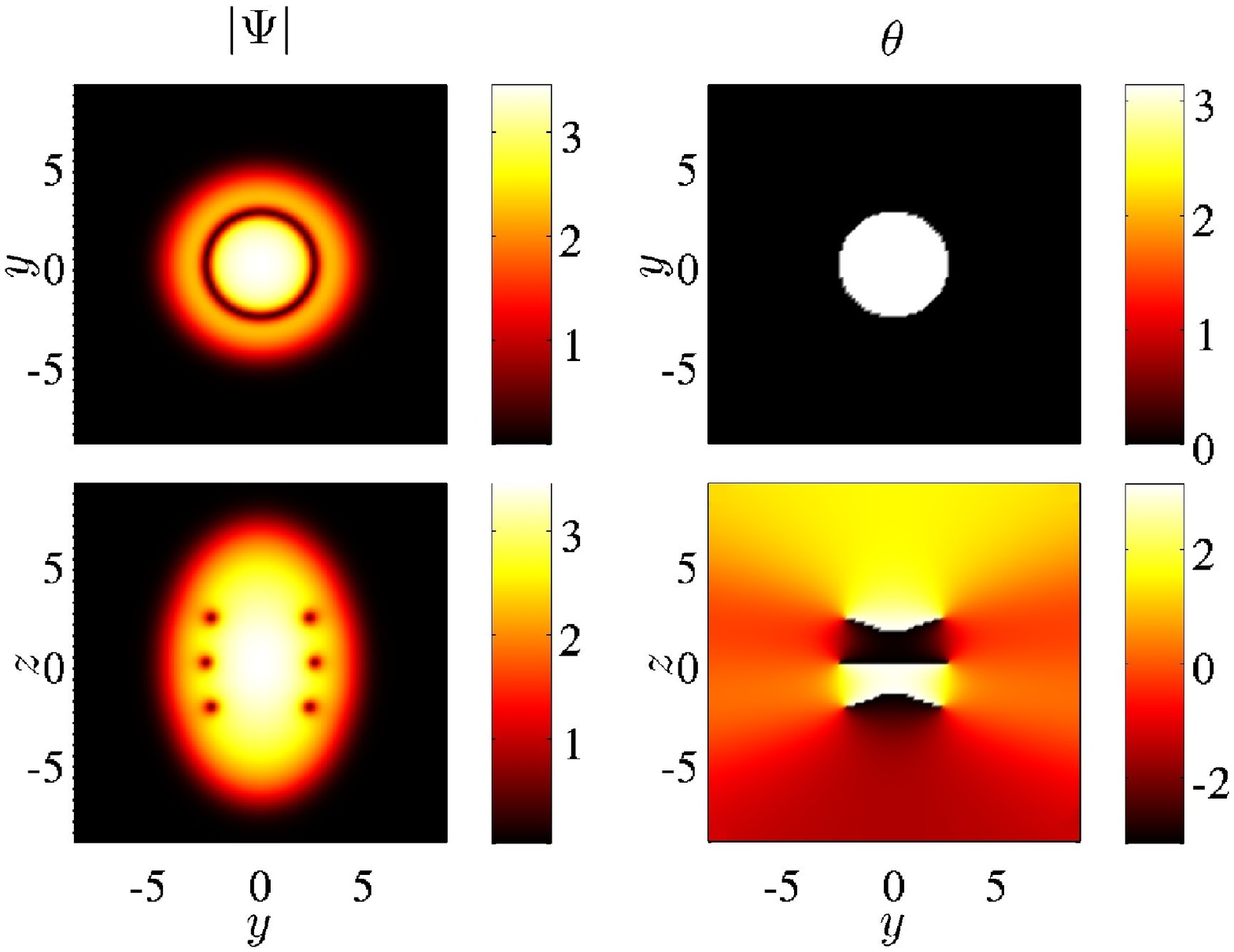}
\includegraphics[width=4.0cm]{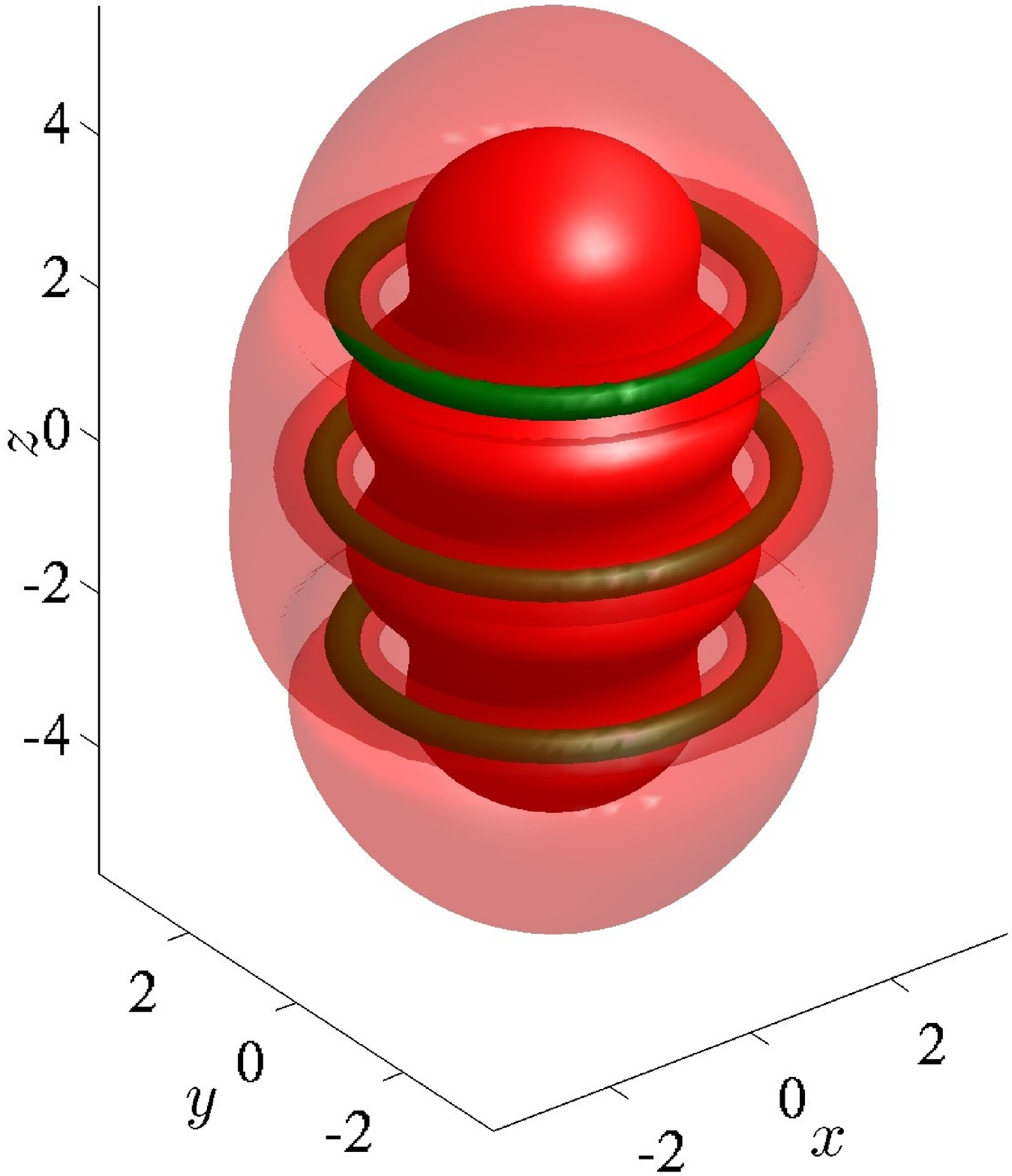}
\includegraphics[width=8cm]{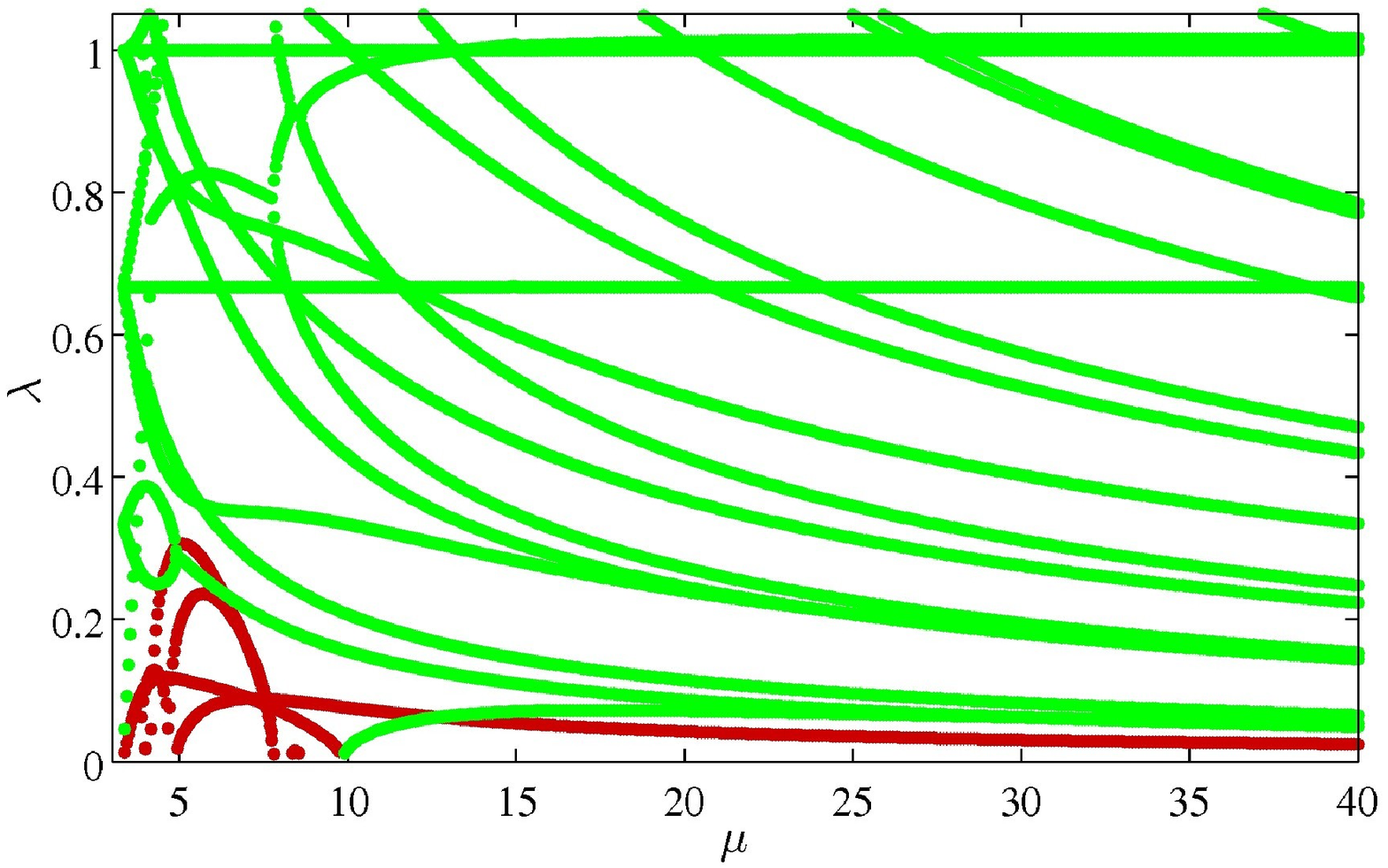}
\caption{(Color online)
The three-VR configuration for $\frac{3}{2}\omega_z=\omega_r=1$.
The four top panels show the modulus $\Psi$ (left panels) and argument $\theta$ 
(right panels) of a triple vortex ring state for a dimensionless 
chemical potential $\mu=12$;
the top row illustrates the $z=0$ plane, while the second row the $x=0$ plane.
The middle panel shows an isocontour density plot.
%
%
The bottom panel depicts the
BdG spectrum for the three-VR as a function
of the dimensionless chemical potential $\mu$.
Same notation as in Fig.~\ref{fig:spectrum2VR}.
%
%
Notice the higher multiplicity of unstable modes and especially the persistent,
albeit weakening, instability due to an eigenmode resulting from the
collision of two modes around $\mu=5$. 
%
}
\label{fig:spectrum3VR}
\end{center}
\end{figure}

\begin{figure}[tb]
\begin{center}
\includegraphics[width=8cm]{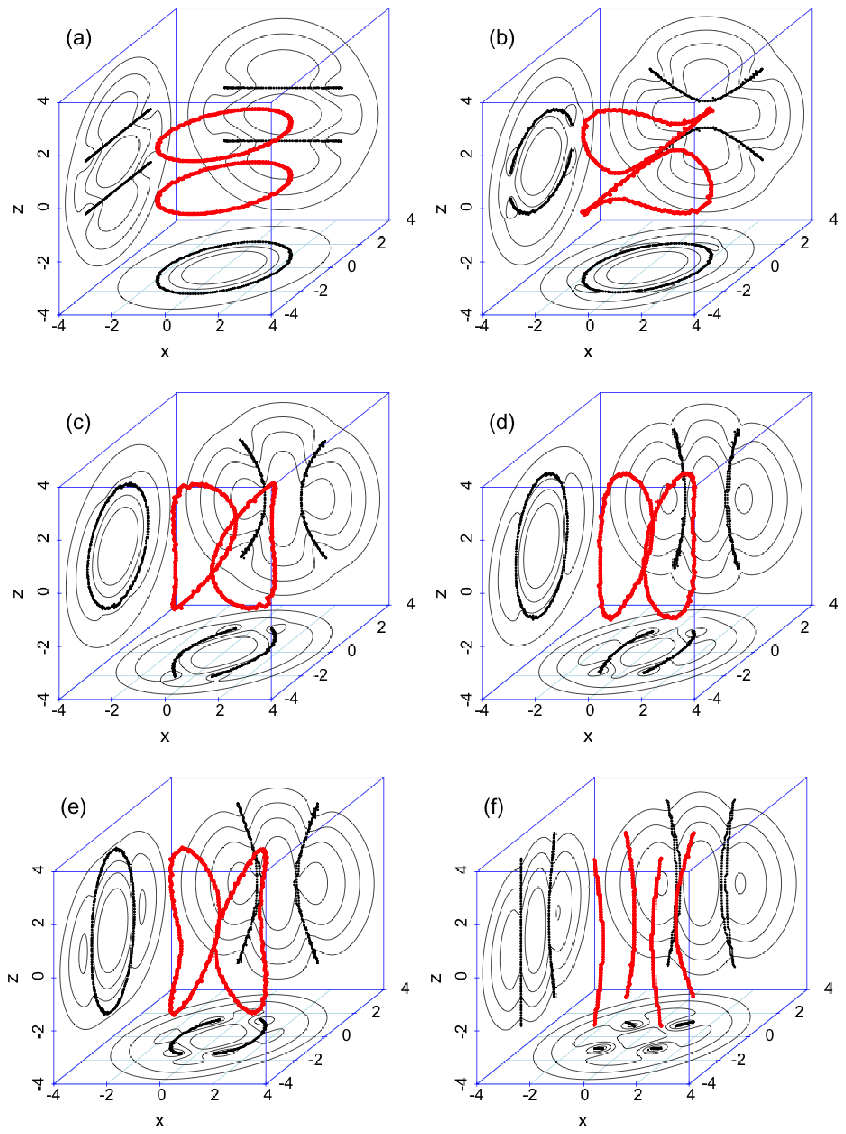} 
\caption{(Color online) 
Evolution of an unstable double vortex ring for chemical potential $\mu=10$.
Six representative snapshots during the time evolution
of the state are given at times
$t$ = (a) 20, (b) 40, (c) 45.4, (d) 68, (e) 71.2 and (f) $74.8$ (units of $1/\omega_r$).
Notice the intense quadrupolar
undulation of the rings leading to their breakup and then reconnection
with a perpendicular axis of symmetry, as well as an example of their
breakup into vortex lines along the $z$-direction. 
}
\label{fig:dyn1}
\end{center}
\end{figure}

\begin{figure}[tb]
\begin{center}
\includegraphics[width=8cm]{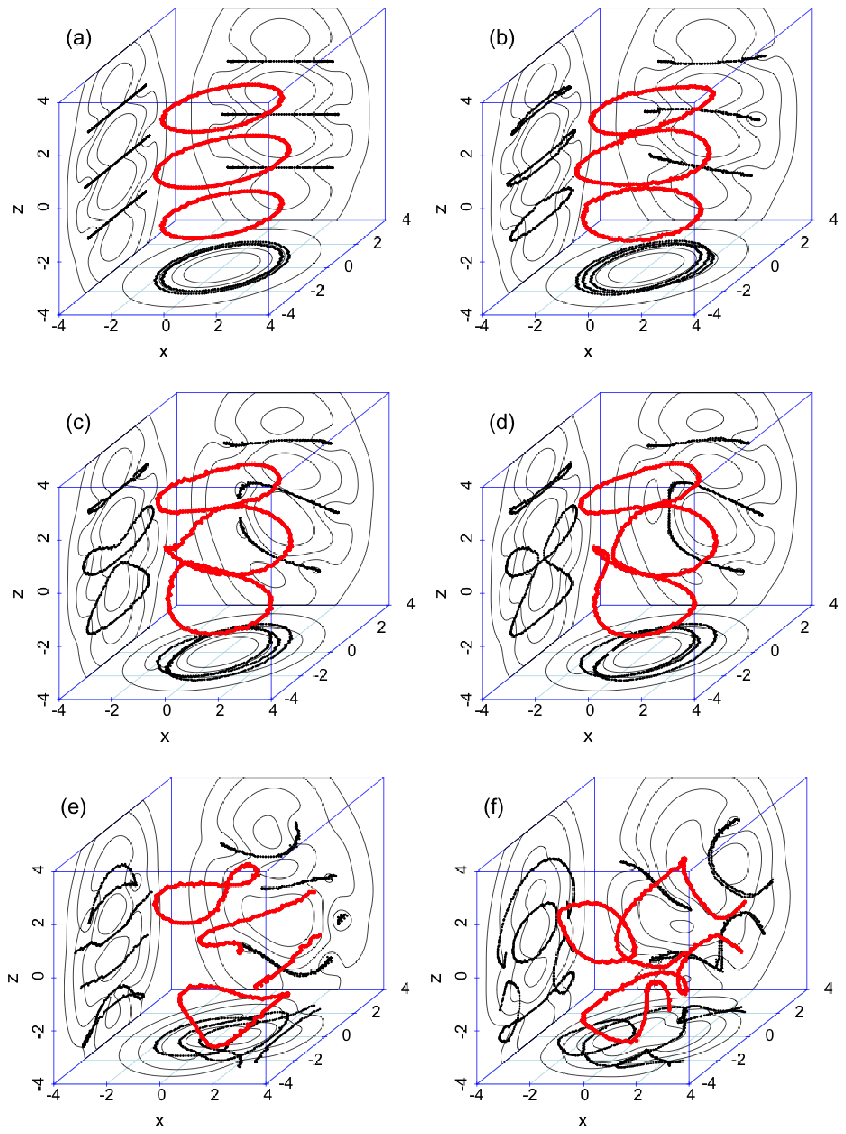} 
\caption{(Color online) 
As in Fig.~\ref{fig:dyn1} but for the 
unstable evolution of the three vortex ring state in the case of $\mu=9$.
The six representative snapshots shown during the time evolution correspond in this case to
$t$ = (a) 20, (b) 49.4, (c) 52.4, (d) 53.3, (e) 54.9 and (f) $56.0$ (units of $1/\omega_r$).
Notice how the deformation of the rings leads, in this case, to the
joining of the two lower ones into a single entity [see (d)], before 
the subsequent evolution to a vortex tangle. 
}
\label{fig:dyn2}
\end{center}
\end{figure}

We also provide a similar spectral perspective in the case of
the three-VR solution in Fig.~\ref{fig:spectrum3VR}. 
%
%
The top quartet and the
middle isocontour panel illustrate a characteristic example of this
state for (dimensionless) chemical potential $\mu=12$.
In the cut
along the central ($x,y$) plane, a ring-like structure is
clearly discernible, while a vertical ($y,z$)-cut reveals
three pairs of opposite-charge vortices, alternating along the
$y$-axis, as well as along the $z$-axis. This implies that
the three rings, given the alternating nature of the constituent triple
soliton pattern, result in an equilibrium triplet
of vortex rings of charge $+1$,
$-1$ and $+1$, up to parity reversals of all three.
The middle
panel clearly showcases, through its density isocontours,
the nature of the pattern.
%
%
%
The bottom panel in the figure depicts the BdG spectrum for the
three-VR solution as the chemical potential is varied.
While some of the oscillatory instabilities associated
with complex eigenvalues appear over narrow parameter
intervals, for small values of $\mu < 10$,
there is an instability
arising from the collision of two eigenmodes near $\mu=5$
that seems to persist for $\mu>10$
and indeed for the entire
parametric interval of chemical potentials that we have explored.
We will explore this dynamical mode in the direct numerical
simulations that follow, as it seems the most pertinent one to the
potential stability of multi-VR states in the TF regime of
large $\mu$. We note that this instability might be \textit{cured} 
by adjusting the trap aspect ratio. Again, we have checked that the 
modes that level off are bulk excitations of the underlying ground state.


We now explore some of the key dynamical features
of multi-VRs through direct numerical simulations.
In Fig.~\ref{fig:dyn1}, we examine a rather intriguing
example of the dynamical instability of the two-VR
state for $\mu=10$.
In this figure we show the 3D positions of the vortex core as red curves \cite{foster} and
their projection onto the $(x,y)$, $(x,z)$ and $(y,z)$ planes are shown as black curves.
2D density contours are also projected on these three planes 
and the contours correspond to 0.25, 0.5, and 0.75 of the maximum density at each time.
%
We observe that the rings initially deform in a quadrupolar
fashion, and once this deformation becomes sufficiently severe,
they split and reconnect as (nearly) co-axial rings perpendicular to their original orientation. The rotated double rings remain robust for a considerable
time interval, as can be seen in frames Fig.~\ref{fig:dyn1}(c)--(e) which span $\Delta t \sim 25$.
We have also observed that, depending on the initial conditions, two VRs can undergo this effective rotation several times, sometimes transiently breaking into four vortex lines, as shown in Fig.~\ref{fig:dyn1}(f), before reforming as a rotated double VR. Eventually, though, the VRs break up into a vortex tangle.

Lastly, we examine a prototypical example of the instability associated
with the three-VR state. We probe, in particular, the unstable mode
that was observed in Fig.~\ref{fig:spectrum3VR} to be the persistent cause
of instability for large values of the chemical
potential. This instability in the case of $\mu=9$
is illustrated in Fig.~\ref{fig:dyn2}. There, it can be observed that the triple VR
becomes subject to a symmetry-breaking deformation of the rings.
As time evolves, the resulting undulations are amplified
and lead to intense Kelvin mode excitations along each
of the rings.
Subsequently, two of these rings may combine, as can be seen in Fig.~\ref{fig:dyn2}(d).
These two rings then separate again, as shown in Fig.~\ref{fig:dyn2}(e), and eventually 
all three rings again evolve into a vortex tangle as shown in Fig.~\ref{fig:dyn2}(f).

\section{Discussion and Future Challenges}
\label{conclusion}

In the present work, we have extended the analytical and numerical
analysis concerning the dynamics of single and multiple VRs
in harmonically confined Bose-Einstein condensates.
We showed that in such systems it is controllably possible
to form and establish the existence of states involving
one-, two-, three- or more VRs, essentially at will
over a wide range of chemical potentials ---or, equivalently,
atom numbers.
Furthermore, we provided a theoretical formulation based 
on the energetics of the different processes involving
the rings at the particle level.
We demonstrated that for the
1-, 2- and 3-VR 
states considered herein this approach,
encompassing the self-induced ring translation, the trap
induced ring oscillation and the inter-ring interaction, could
capture qualitatively and in some cases quantitatively
the ring steady states.
%
The stability for these multi-VR steady states was also probed at the level 
of a full Bogoliubov-de Gennes analysis which revealed both the instabilities
of, e.g., 2- and 3-VR states, but also the collective relative motions
of the rings.
Finally, some selected examples of the dynamics of the VRs
were displayed, demonstrating both their potential for coherent
multi-VR motions, and also their rich instability scenarios
containing examples of VR breakups and recombinations.
Such events include ring mergers and subsequent splits or redistribution 
of the vorticity in the form of vortex lines.

There are numerous questions that remain
open and constitute interesting directions for future work.

First, in the context of a single vortex ring, identifying
an approach, perhaps involving an adiabatic invariant
as in Refs.~\cite{ba,kono,kamch,ai}), that would
give a more accurate description of its equilibrium
and overall dynamics would be helpful both in the
realm of single and in that of multiple VRs.

{%
Secondly, regarding the case of more than one VR, 
it is important to mention that, although the PP
gave a qualitative description of the size and location 
of the multiple VR steady-state configuration, it fails 
to accurately describe the precise location 
of the fixed points in a quantitative fashion.
Nonetheless, it is interesting to note that, despite this shortcoming,
the reduced PP equations of motion are able to 
capture (results not shown here):
(a) all the different modes of vibration of the VRs (for two VRs the
in-phase and our-of-phase oscillations and for three VRs the three
normal modes of vibration)
and (b) the trend corresponding to the actual frequency of
vibration for these oscillatory vibrational modes as the
chemical potential in varied.
}

In the same vein, it would be interesting, albeit
technically demanding, to attempt to incorporate
the density modulations in the inter-ring interactions,
in a similar way to what was proposed for ordinary vortices
in Ref.~\cite{busch}.
A perhaps more straightforward,
although computationally demanding possibility, would
be to explore the dynamics of VRs in a way similar to that of
Ref.~\cite{baggaley}. In this work, the
authors use the full Biot-Savart law in order to explore
the interactions between numerous initially co-axial
vortex rings in a uniform medium with the aim of
studying their leapfrogging behavior. The Biot-Savart
law encompasses two out of the three features we are
considering here, namely the ring self-action and
the inter-ring interaction. It does not include the
effect of the trap, yet this can be accounted for through the
work of Ref.~\cite{fetterpra}. 
The combination of these approaches would then enable at an effective, 
rather than full GPE, level the examination
of not solely co-axial vortex rings as is done here but
rather allowing their deformation (beyond co-axiality)
while considering their motion. Comparing
such a sophisticated approach with the full Gross-Pitaevskii 
dynamics would be especially informative.

Naturally, also, extensions of the present considerations
to a higher number of components, including the potential
formation of more exotic structural configurations such
as Skyrmions~\cite{ruost1,ruost2} would be of interest
as well for future work.

Progress along some of these directions, as relevant, will be reported 
in future publications.

\begin{appendix}
\section{Effective ODEs for trapped VRs}
\label{sub:append}

Here we include the explicit effective  ODEs describing
the dynamics of $N$ VRs with radii $r_i$ and vertical positions $z_i$
inside an isotropic trap with strength $\omega$
and TF radius $R_\perp=\sqrt{2\mu}/\omega$ for chemical potential $\mu$.
By using Hamilton's equations (\ref{eq:HJ}), the dynamics for each
VR is given by:
\begin{eqnarray}
\dot r_i &=&  
\dot r_i^{(\text{VR-T})} +
\dot r_i^{(\text{VR-VR})} ,
\notag
\\
\notag
\dot z_i &=& 
\dot z_i^{(\text{VR-T})} + 
\dot z_i^{(\text{VR-VR})} ,
\end{eqnarray}
where the contributions due to the trap are encapsulated in the
VR-T terms and the vortex-vortex interactions in the VR-VR terms.
The VR-T contributions yield
\begin{eqnarray}
\dot r_i^{(\text{VR-T})} &=& 2\alpha\, r_i z_i (L-1),
\notag
\\
\notag
\dot z_i^{(\text{VR-T})} &=& \alpha(R_\perp^{2}-3\,r_i^{2}-z_i^{2}) L 
-2\alpha(R_\perp^{2}-2\,r_i^{2}-z_i^{2}) ,
\end{eqnarray}
where
\begin{eqnarray}
L&=& \ln  \left( {
\frac {{R_\perp^{2}-(r_i^{2}+z_i^{2})}}{\xi^2}} \right),
\notag
\\
\notag
\alpha &=& \frac {{\pi }^{2}\mu}{2 r_i R_\perp^{2}}.
\end{eqnarray}
On the other hand, the VR-VR contributions yield
\begin{eqnarray}
\dot r_i^{(\text{VR-VR})} &=& 
-\frac{1}{m_i r_i} \sum_{{j\neq i}}^N \frac{\partial W_{ij}}{\partial z_i}
\notag
\\
\notag
\dot z_i^{(\text{VR-VR})} &=& 
+ \dfrac{1}{m_i r_i} \sum_{{j\neq i}}^N \frac{\partial W_{ij}}{\partial r_i}
\label{riziK}
\end{eqnarray}
where
\begin{eqnarray}
\notag
W_{ij} &=& 
 m_i m_j\, \sqrt{r_i r_j}\,C(k_{ij}),
\end{eqnarray}
and $C(k_{ij})$ is defined in Eq.~(\ref{eq:Cij}).
%

\end{appendix}

\begin{acknowledgments}

W.W.~acknowledges support from NSF-DMR-1151387 and from the Office of the Director of 
National Intelligence (ODNI), Intelligence Advance Research Projects Activity (IARPA),
via MIT Lincoln Laboratory Air Force Contract No.~FA8721-05-C-0002.
R.N.B. is supported by the QUIC grant of the Horizon2020 FET program and by 
Provincia Autonoma di Trento.
R.N.B., C.T., L.A.C. and P.G.K. acknowledge support by Los Alamos
National Laboratory, which is operated by LANS, LLC
for the NNSA of the U.S.~DOE and, specifically, Contract No. DEAC52-06NA25396.
R.C.G.~gratefully acknowledges the support of NSF-DMS-1309035 and PHY-1603058. 
P.G.K.~gratefully acknowledges the support of NSF-DMS-1312856 and PHY-1602994, 
from the ERC under FP7, Marie Curie Actions, People, 
International Research Staff Exchange Scheme (IRSES-605096),
and the Stavros Niarchos Foundation via the Greek Diaspora Fellowship Program.
The views and conclusions contained herein are those of the authors and
should not be interpreted as necessarily representing the official policies
or endorsements, either expressed or implied, of ODNI, IARPA, or the
U.S.~Government.
The U.S.~Government is authorized to reproduce and distribute reprints for 
Governmental purpose notwithstanding any copyright annotation thereon.
We thank the Texas A\&M University for access to their Ada cluster.

\end{acknowledgments}


\begin{thebibliography}{99}
\bibitem{dvr-exp}  As an example, for the GPE solution of the two-vortex 
ring spherical case with a chemical potential $\sim 16\hbar\omega_r$ and 
$4\pi Na_s\sqrt{m\omega_r/\hbar}$ = 4413, we employ  in each Cartesian 
coordinate a grid  of 257 points derived from 64 elements with an order 
5 Gauss-Legendre quadrature, spanning a range of  $[-8.2,8.2]$ trap units, 
with a time step of $1.0\times10^{-4}$].
\end{thebibliography}
\end{document}